\documentclass[twocolumn,prl,superscriptaddress,bibnotes,notitlepage,nofootinbib]{revtex4-1}
\usepackage{color}
\definecolor{red}{rgb}{0.75,0,0}
\definecolor{blue}{rgb}{0,0,0.75}
\definecolor{green}{rgb}{0,0.5,0}
\usepackage{amsmath}
\usepackage{amssymb}
\usepackage{lipsum}
\usepackage{bm}
\usepackage{xcolor}
\usepackage{graphicx}
\usepackage{verbatim}
\usepackage[T1]{fontenc}
\usepackage{subcaption}
\usepackage{hyperref}

\def\be{\begin{equation}}
\def\ee{\end{equation}}
\def\bea{\begin{eqnarray}}
\def\eea{\end{eqnarray}}

\def\besub{\begin{subequations}}
\def\eesub{\end{subequations}}

\def\bwd{\begin{widetext}}
\def\ewd{\end{widetext}}

\definecolor{ao(english)}{rgb}{0.0, 0.5, 0.0}
\definecolor{armygreen}{rgb}{0.29, 0.33, 0.13}
\definecolor{auburn}{rgb}{0.43, 0.21, 0.1}
\definecolor{brightmaroon}{rgb}{0.76, 0.13, 0.28}
\definecolor{cadmiumred}{rgb}{0.89, 0.0, 0.13}
\definecolor{carnelian}{rgb}{0.7, 0.11, 0.11}
\definecolor{cornellred}{rgb}{0.7, 0.11, 0.11}
\definecolor{crimsonglory}{rgb}{0.75, 0.0, 0.2}
\definecolor{orangeyellow}{rgb}{0.3, 0.2, 0.2}
\definecolor{fluorescentorange}{rgb}{1.0, 0.75, 0.0}
\definecolor{gamboge}{rgb}{0.89, 0.61, 0.06}
\newcommand{\bsf}[1]{\textsf{\textbf{#1}}}

\newcommand{\AM}[1]{\textcolor{black}{#1}}

\newcommand{\AMR}[1]{\textcolor{black}{#1}}

\begin{document}
\title{Chiral active hexatics: Giant number fluctuations, waves and destruction of order}
\author{Ananyo Maitra}
\email{nyomaitra07@gmail.com}
\affiliation{Sorbonne Universit\'{e} and CNRS, Laboratoire Jean Perrin, F-75005, Paris, France}
\author{Martin Lenz}
\affiliation{LPTMS, CNRS, Univ. Paris-Sud, Universit\'e Paris-Saclay, 91405 Orsay, France}
\affiliation{PMMH, CNRS, ESPCI Paris, PSL University, Sorbonne Universit\'{e},
Universit\'{e} de Paris, F-75005, Paris, France}
\author{Raphael Voituriez}
\affiliation{Sorbonne Universit\'{e} and CNRS, Laboratoire Jean Perrin, F-75005, Paris, France}
\affiliation{Sorbonne Universit\'{e} and CNRS, Laboratoire de Physique Th\'{e}orique de la Mati\`{e}re Condens\'{e}e,
F-75005, Paris, France}

\begin{abstract}
Active materials, composed of internally driven particles, have properties that are qualitatively distinct from matter at thermal equilibrium. However, the most spectacular departures from equilibrium phase behaviour are thought to be confined to systems with polar or nematic asymmetry. In this paper we show that such departures are also displayed in more symmetric phases such as hexatics if in addition the constituent particles have chiral asymmetry. We show that chiral active hexatics whose rotation rate does not depend on density have giant number fluctuations. If the rotation rate depends on density, the giant number fluctuations are suppressed due to a novel orientation-density sound mode with a linear dispersion which propagates even in the overdamped limit. However, we demonstrate that beyond a finite but large lengthscale, a chirality and activity-induced relevant nonlinearity invalidates the predictions of the linear theory and destroys the hexatic order. In addition, we show that activity modifies the interactions between defects in the active chiral hexatic phase, making them non-mutual. Finally, to demonstrate the generality of a chiral active hexatic phase we show that it results from the melting of chiral active crystals in finite systems.
\end{abstract}

\maketitle
Active matter is driven out of equilibrium by a continuous supply of energy at a microscopic scale which leads to macroscopic forces and currents \cite{SR JSTAT, LPDJSTAT, RMP}. Continuum active hydrodynamic theories \cite{TonTuRam, Prost_nat} for such nonequilibrium states have been constructed for multiple active liquid-crystalline phases and have, more recently, been extended to include chiral asymmetry \cite{Lewis1, Lewis2, Strempel, seb3, Cates_drop}. 
The interaction between chirality and activity, especially in two dimensions, leads to several surprising features including the suppression of the generic instability \cite{Aditi1, RMP} in orientationally-ordered active fluids \cite{Ano_chi}, odd viscosity waves in isotropic chiral fluids \cite{Deboo1} and waves in overdamped  chiral active solids \cite{Deboo2}.

In this paper we consider large scale properties of \emph{hexatic} chiral active systems in two dimensions. This phase can arise upon dislocation unbinding of chiral active solids \cite{Deboo2} just as passive hexatic ordered phases result from the solid phase. Chiral hexatic phases can also be realised experimentally in artificial active systems in which the elementary units self assembly into chiral structures with hexagonal symmetry \cite{Aubret}. 
Further, hexatic correlations are observed in cell-layers and tissues \cite{Hexagonal Packing of Drosophila Wing Epithelial Cells by the Planar Cell Polarity Pathway, The mechanical anisotropy in a tissue promotes ordering in hexagonal cell packing} and since multiple cell lines are chiral \cite{Spontaneous shear flow in confined cellular nematics} they should be described by our theory. A chiral hexagonal phase has also been observed in spermatozoa at a planar interface \cite{Kruse} and bacteria has been shown to organise into a chiral hexagonal crystal \cite{Fast-Moving Bacteria Self-Organize into Active Two-Dimensional Crystals of Rotating Cells}. Furthermore, hexagonal organisation of chiral microscopic units is common even at the subcellular level, for instance in clathrin coats \cite{MartinL}. The hydrodynamic theory we construct also describes \emph{all} $n-$atic chiral active phases, for $n>2$ and our results are valid for \emph{all} of them, including tetratics.
Finally our theory may be more widely applicable even in systems in which microscopic constituents are not themselves chiral but in which the chiral symmetry is spontaneously broken \cite{palacci}.

Much of the interesting phase behaviour in active systems arise from the interaction of the nonequilibrium drive with dipolar or quadrupolar asymmetry. This leads to ``sound modes'' with linear dispersion in \emph{overdamped} polar systems \cite{Toner_Tu} and giant number fluctuations in polar and nematic systems \cite{Toner_Tu, Aditi2, Ano_apol,RMP}. However, it is believed that more symmetric phases do not display such spectacular departures from equilibrium behaviour \cite{Cugliandolo, Ano_apol}. This is broadly correct for \emph{achiral} hexatic phases. However, we demonstrate that this is \emph{not} correct for \emph{chiral} hexatic phases which may display autonomous rotation.  We show that the \AMR{chiral} active hexatic phase has anomalous number fluctuations  -- the root-mean-squared number fluctuations $\sqrt{\delta N^2}$ in a region containing on average $N$ particles scaling as $N$ instead of $\sqrt{N}$ as in equilibrium -- when the global rotation rate is constant. However, a \emph{density-dependent} rotation rate changes the picture significantly: the giant number fluctuations are suppressed and the coupled density-orientational fluctuations lead to either a wave with a linear dispersion relation or an instability with a linear growth rate at small wavevectors even in the overdamped limit.
However, the interplay of activity and chirality yields a \emph{relevant} nonlinearity leading to the destruction of the quasi-long-range-ordered (QLRO) hexatic state in two dimensions and implying that the earlier predictions only applies for systems below a critical size. We then demonstrate that the behaviour of both single topological defects \emph{and} interaction between defects are modified by chirality and activity. A single defect spontaneously rotates and its angular far-field is modified due to the interplay of activity and chirality.
The force between two oppositely-charged defects is also modified and becomes non-mutual, \AMR{ i.e., the forces between a pair of defects are not equal and opposite}, leading to a directed motion of a defect-pair on a substrate. \AMR{Non-mutual interactions between defects should be a general, but hitherto unstudied, feature of \emph{all} active orientationally ordered systems irrespective of whether they are chiral or not.}  Finally, we will discuss how an active chiral hexatic emerges from a chiral solid and relate the phenomenological coefficients in the chiral hexatic to those in the active solid.

We first discuss non-rotating chiral hexatics on a substrate \AMR{and construct its equation of motion including all terms allowed by symmetry to lowest order in gradients and fields \cite{MPP, ZHN}}. The local density of the particles is described  by $\rho({\bf x}, t)$, which obeys a continuity equation $\partial_t\rho=-\nabla\cdot(\rho{\bf v})$ where ${\bf v}({\bf x}, t)$ is their velocity field.
The six-fold-symmetric phase is characterised by the complex order parameter $\Psi=\psi e^{6i\theta({\bf x}, t)}$ where, for a phase with bond-angle order, $\theta({\bf x},t)$ denotes the orientation of the line joining two neighbouring particles with respect to an arbitrary but fixed axis. It can also describe the orientation of more complex elementary \AMR{or self-assembled} units that are themselves six-fold symmetric \cite{Aubret, MartinL}. \AMR{As discussed in greater detail in \cite{supp}, the dynamics of small angular deviations about an ordered hexatic is }
\begin{equation}
\label{bang}
\AMR{\frac{d{\theta}}{dt}}=\frac{1}{2}\epsilon_{ij}\partial_i v_j+\gamma_c\nabla\cdot{\bf v}-\Gamma_6\frac{\delta F}{\delta\theta}+\xi_\theta({\bf x}, t)
\end{equation}
\AMR{where $d/dt$ denotes a convected derivative $\partial_t+{\bf v}\cdot\nabla$}, $F=\int d{\bf r}K(\nabla\theta)^2/2+f(\rho)$ is the free energy with $f(\rho)$ being a function of density, $\Gamma_6$ is a dissipative kinetic coefficient, $\xi_\theta({\bf x}, t)$ is a Gaussian white noise of strength $\Delta_\theta$ and $\boldsymbol{\epsilon}$ 
is the two-dimensional Levi-Civita tensor.
The first term on the R.H.S. of \eqref{bang} denotes the precession of the angular distortion in a local vorticity field \cite{ZHN}. The second is a chiral \AMR{but \emph{passive} reversible coupling between $\theta$ and ${\bf v}$ and} leads to a chiral precession in response to a local isotropic compression or dilation. 
\begin{figure}
  \centering
  \includegraphics[width=7.5cm]{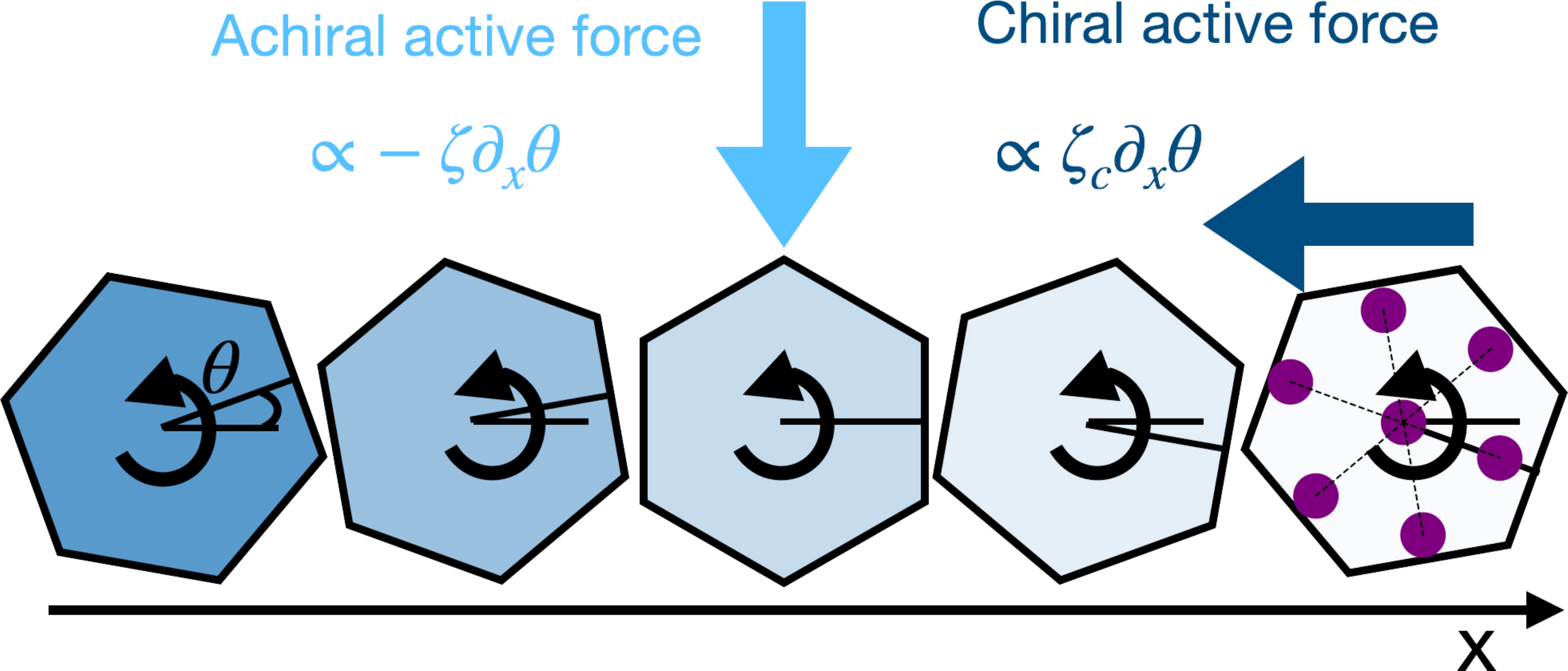}
\caption{\AMR{Illustration of the hexatic active forces. The hexagons represent either hexagonal particles or clusters of particles with hexatic order (we represent such a cluster inside the rightmost hexagon). The circular arrows denote the handedness and the colour gradient of the hexagons denote the angular gradient of the hexatic field. The directions of the chiral and achiral active forces are shown in the figure for a distortion of along $\hat{x}$.}}
\label{chi_for_fig}
\end{figure}
The third term controls the relaxation of the bond-angle order parameter to its equilibrium value in the absence of activity and flow. Finally \AMR{\cite{supp}}, the equation for the velocity field is 
\begin{multline}
\label{velhex}
\AMR{\rho\frac{d{{\bf v}}}{dt}}=-\Gamma{\bf v}-{\gamma_cK}\nabla\nabla^2\theta-\frac{K}{2}(\hat{z}\times\nabla)\nabla^2\theta-\rho\nabla\frac{\delta F}{\delta\rho}\\-\zeta\hat{z}\times\nabla\theta-\zeta_c\hat{z}\times(\hat{z}\times\nabla)\theta+\Gamma_c\hat{z}\times{\bf v}+\boldsymbol{\xi}_v
\end{multline}
The terms in the first line of \eqref{velhex} are passive forces, with the first being the usual friction and the others being passive couplings to the orientation and density fields. The first three terms in the second line are active while the last is a Gaussian white noise of strength $\Delta_v$. The first active term, with the coefficient $\zeta$, is an achiral force equivalent to the one discussed in \cite{Ano_apol}. The second, which can be rewritten as $\zeta_c\nabla\theta$, is explicitly chiral, with the handedness being encoded in the sign of $\zeta_c$ (see Fig. \ref{chi_for_fig}). The final term is an \emph{active chiral} friction. \AMR{At this order in gradients and fields, \eqref{bang} and \eqref{velhex} can describe \emph{any} $n-$atic phase for $n>2$. However, they \emph{cannot} describe generic nematic and polar chiral phases which feature additional active \emph{and} passive terms at this order \cite{Ano_chi} forbidden by $n-$fold symmetry for $n>2$.}

We eliminate ${\bf v}$ using \eqref{velhex} in the overdamped limit to obtain coupled linearised hydrodynamic equations for the density and angle fields in Fourier space:
\begin{multline}
\label{bangfluc}
\partial_t\theta=-q^2\Bigg[\left(\Gamma_6 K+\frac{\Gamma_c\zeta_c-\Gamma\zeta+2\gamma_c(\Gamma\zeta_c+\zeta\Gamma_c)}{2(\Gamma^2+\Gamma_c^2)}\right)\theta\\-\frac{\Gamma_c+2\Gamma\gamma_c}{2(\Gamma^2+\Gamma_c^2)}\rho\Bigg]+\xi_\theta
\end{multline}
\begin{equation}
\label{densfluc}
\partial_t\rho=-q^2\rho_0\left[-\frac{\Gamma\zeta_c+\zeta\Gamma_c}{\Gamma^2+\Gamma_c^2}\theta+\frac{\Gamma}{\Gamma^2+\Gamma_c^2}\rho\right]+\xi_\rho.
\end{equation}
where $\xi_\rho$ is a conserving Gaussian white noise inherited from \eqref{velhex} and has the correlation $q^2\Delta_v/(\Gamma^2+\Gamma_c^2)$ \AMR{and $\rho_0$ is the steady state density}.
In achiral but active hexatics, the $\theta$ and $\rho$ equations would have been linearly decoupled at this order in wavevectors. The coupling to the angle field in \eqref{densfluc}, which is purely chiral,  yields a mass density current $\propto i{\bf q}\theta$ which in the steady-state must be balanced by diffusive current $\propto i{\bf q}\rho$ \cite{SR_rev}. Therefore, density fluctuations in chiral active hexatics must scale as orientational fluctuations just as in active nematics. The orientational fluctuations are Goldstone modes of broken rotational symmetry and $\lim_{q\to 0}\langle|\theta|^2\rangle \sim 1/q^2$ \cite{Forster} and as we explicitly show in \cite{supp}; this implies that the static structure factor of density fluctuations is $\lim_{q\to 0}\langle|\delta\rho|^2\rangle \sim 1/q^2$. This is in contrast to passive systems and achiral active hexatics in which it goes to a constant as $q\to 0$. The number fluctuations are, therefore, giant -- a region containing on average $N$ particles must have R.M.S. fluctuations $\sqrt{\delta N^2}\sim N$ -- which was hitherto believed to require polar or nematic asymmetry \cite{Cugliandolo, Ano_apol}. This demonstrates that \emph{all} $n-$atic systems, including tetratics and hexatics should display giant number fluctuations \AMR{provided} they are chiral. 

\AMR{We now consider a purely chiral nonequilibrium phase in which the bond angle field rotates \emph{locally} at a rate $\Omega$ i.e. $\Psi ({\bf x}, t)=\psi e^{6i(\theta-\Omega t)}$. This autonomous rotation of the particles is distinct from a global rotation of the system.}
\begin{figure}
  \centering
  \includegraphics[width=9cm]{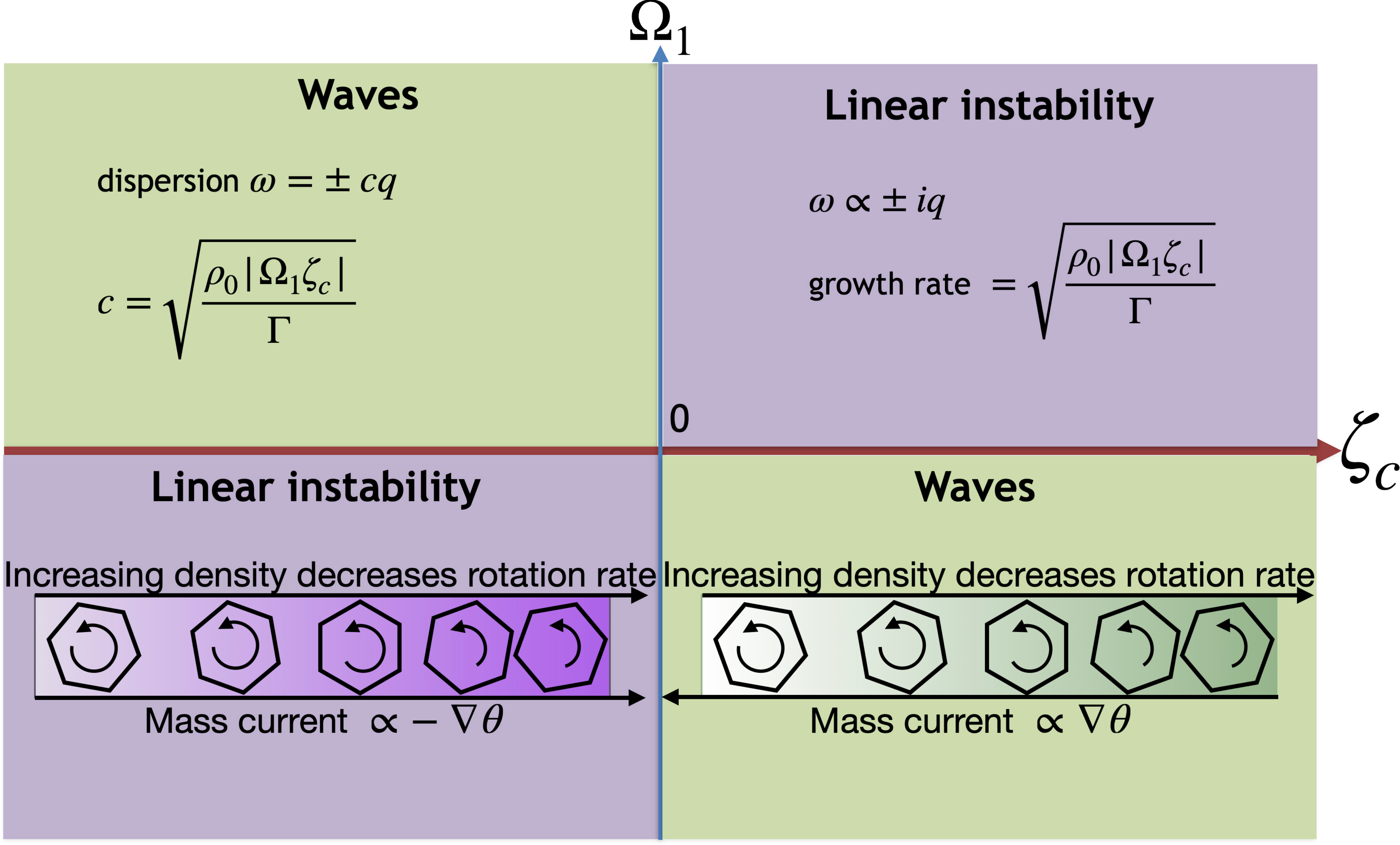}
\caption{\AMR{Summary of the linear theory of chiral hexatics for the simple case $\Gamma_c=0$. For $\Omega_1\zeta_c<0$ there are propagating wave modes with a linear dispersion while for $\Omega_1\zeta_c>0$, the chiral hexatic phase is unstable with a growth rate linear in wavenumber. The chiral hexatic displays giant number fluctuations only along the $\Omega_1=0$ line and normal number fluctuations elsewhere. The colour gradient represents density variation and the circular arrows represent rotation with the speed of rotation being represented by the length of the arrows.}}
\label{rotation_instab}
\end{figure}
This implies that $\Omega$ must be added to the R.H.S. of \eqref{bang}, where 
\begin{equation}
\Omega(\rho)=\Omega(\rho)|_{\rho=\rho_0}+\frac{\partial\Omega}{\partial\rho}\bigg|_{\rho=\rho_0}\delta\rho=\Omega_0+\Omega_1\delta\rho.
\end{equation}
The constant rotation rate can be eliminated by \AMR{a simple change of variable} $\theta\to\theta-\Omega_0 t$ implying that the results discussed for a non-rotating active chiral hexatic also apply to its steadily rotating ($\Omega_1=0$) counterpart. However, a density-dependent rotation rate ($\Omega_1\neq 0$) fundamentally modifies the phase behaviour.
The lowest order in wavevector equation for $\theta$ is 
$\partial_t\theta= \Omega_1\delta\rho+\mathcal{O}(q^2)$ while the density dynamics is still described by \eqref{densfluc}. This yields a \emph{linear} dispersion $\omega_\pm=\pm c q$ where 
\begin{equation}
\label{wavesp}
c=\sqrt{-\frac{\rho_0\Omega_1(\Gamma_c\zeta+\Gamma\zeta_c)}{\Gamma^2+\Gamma_c^2}}.
\end{equation}
When $\Omega_1(\zeta_c\Gamma+\Gamma_c\zeta)<0$, i.e. either when the rotation slows down with increasing density while particles tend to diffuse \emph{up} gradients of $\theta$ or when the rotation speeds up while particles diffuse down gradients of $\theta$, this leads to a propagating density-orientation wave with a \emph{linear dispersion} (Fig. \ref{rotation_instab}). Such a linear sound-wave-like mode in an overdamped system is only possible in an active system and was previously thought to require polarity \cite{Toner_Tu, SR_rev, RMP}. Here, however, they arise due to \emph{chiral} asymmetry. This chiral active current also reduces the number fluctuations which now obeys the law of large numbers: $\lim_{q\to 0}\langle|\delta\rho|^2\rangle\sim \text{const.}$ as shown in \cite{supp}. Heuristically, since $\omega\sim q$ and $-i\omega \delta\rho\sim -q^2\theta$, $\delta\rho\sim - i q\theta$ and therefore $\langle|\delta\rho|^2\rangle\sim q^2\langle|\theta|^2\rangle\sim q^2(1/q^2)$.
Eq. \eqref{wavesp} further implies that when $\Omega_1(\zeta_c\Gamma+\Gamma_c\zeta)>0$, the homogeneous hexatic phase is \emph{unstable} with a growth rate of fluctuations $\propto q$ (see Fig. \ref{rotation_instab}). If  $\Omega_1>0$ and $(\zeta_c\Gamma+\Gamma_c\zeta)>0$, an angular gradient leads to a mass current \emph{in the direction of the gradient} leading to an increase in density. This higher density leads to a local increase of the rotation rate reinforcing the angular gradient and leading to an instability of the homogeneous hexatic state, possibly towards a patterned structure.

We have till now considered only linear deviations away from the steady-state. We now check whether nonlinear terms affect the conclusions reached using the linear theory. For this, we first consider a ``Malthusian'' hexatic \cite{Malthusian} i.e, one in which the density is not globally conserved, but locally held fixed. The nonlinearity with fewest gradients and fields in the equation for angular fluctuations is $\propto (\nabla\theta)^2$ which arises from the advective nonlinearity ${\bf v}\cdot\nabla\theta$ in \eqref{bang}, since ${\bf v}$ has a chiral contribution $\propto \nabla\theta$. The nonlinear equation of motion for the angle field of a noisy Malthusian hexatic, upon eliminating the velocity field is
\begin{equation}
\label{comKPZ}
\partial_t\theta=\frac{\lambda}{2}(\nabla\theta)^2+\bar{K}\nabla^2\theta+\xi_\theta
\end{equation}
where $\lambda=-2{\Gamma\zeta_c+\zeta\Gamma_c}/({\Gamma^2+\Gamma_c^2})$ and
\begin{equation}
\bar{K}=\left(\Gamma_6 K+\frac{\Gamma_c\zeta_c-\Gamma\zeta+2\gamma_c(\Gamma\zeta_c+\zeta\Gamma_c)}{2(\Gamma^2+\Gamma_c^2)}\right)
\end{equation}
in terms of the previously introduced variables.
Eq. \eqref{comKPZ} has the same form as the KPZ equation with the only distinction being that $\theta$ is a periodic variable. The nonlinearity with the coefficient $\lambda$ is \emph{marginally relevant} in two dimensions i.e., it grows larger upon renormalisation, invalidating the linear theory at large scales \cite{KPZorig, TonerPRX}. This yields the ``rough state'' at large scales with an algebraic scaling of height fluctuations. In the context of the hexatic phase this implies that the hexatic state loses even algebraic order beyond the scale $L_*\sim e^{{16\pi\bar{K}^3}/{\Delta_\theta\lambda^2}}$ \cite{TonerPRX, Diehl1}. While this calculation is for a Malthusian hexatic system, a coupling to the density field cannot make the angular fluctuations \emph{less} divergent. Indeed, the lowest order nonlinearity in the angle field equation that couples density and angular fluctuations has the form $\nabla\delta\rho\cdot\nabla\theta$. Since the linear static structure factor of the density fluctuations is either as large as angular fluctuations (for a density-independent rotation rate) or smaller (for a density dependent rotation rate), this nonlinearity is either as relevant as $(\nabla\theta)^2$ or less relevant than it. In either case, the conclusion that even algebraic order is destroyed due to the $(\nabla\theta)^2$ nonlinearity cannot be modified by the extra nonlinearities coupling density and angular fluctuations.

\begin{figure}
  \centering
  \includegraphics[width=8.5cm]{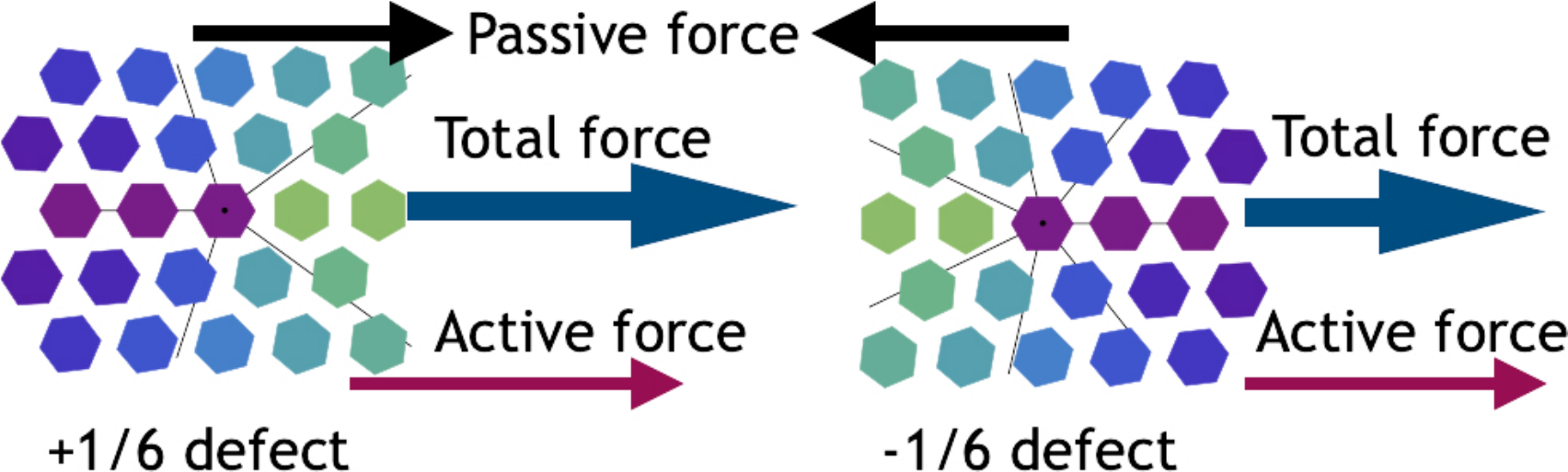}
\caption{\AM{Non-mutual interaction between $\pm 1/6$ defects. The passive interaction between the defect-pair is attractive and mutual, but the active force is non-mutual. Therefore, a $+1/6$ defect may be attracted to a $-1/6$ defect while repelling the $-1/6$ defect leading to the net motion of the $\pm 1/6$ pair. In addition to this, the hexatic fields around $+1/6$ and $-1/6$ defects also spontaneously rotate in \emph{opposite} senses.}}
\label{hex_def}
\end{figure}
We have, till now, only considered smooth fluctuations of the angle field. While we have already demonstrated that the chiral hexatic only has short-range order, topological defects, which in a hexatic predominantly have charges $\pm 1/6$, can still have significant impact on the phase behaviour. In \cite{supp}, we show that chirality leads to a spontaneous rotation of a single defect with an angular speed $\propto \zeta_c$ \emph{even when} a defect-free hexatic does not rotate. The $\lambda$ term in \eqref{comKPZ} also leads to a qualitative modification of the far field structure of the angle-field due to a defect -- unlike in achiral hexatics, in which the angular far-field is independent of the distance from the defect core, in chiral active hexatics it is an explicit function of this. This is similar to defect structures in the, so-called, compact KPZ equation \cite{Diehl1} and spiral waves in complex Ginzburg-Landau equation. Activity also qualitatively modifies the interaction between two defects. In particular, we show in \cite{supp} that the interactions between defects are \emph{non-mutual} -- the strength of attractive or repulsive force exerted on a $+1/6$ defect by a $-1/6$ defect is different from the force exerted by a $+1/6$ defect on a $-1/6$ --  implying that though a single $+1/6$ or $-1/6$ defect doesn't self-propel \AMR{due to the symmetry of the angle field around them}, a $\pm 1/6$ pair, maintained at a fixed separation, does (see Fig. \ref{hex_def}). \AMR{Moreover, while we explicitly demonstrate the non-mutual interaction between defects in the case of chiral active hexatics, the interaction between defects in \emph{all} chiral or achiral active orientationally ordered phases, including nematic and polar systems, should be non-mutual. Defect \emph{pairs} in all such systems should self-propel irrespective of whether single defects self-propel or not.}
Further, \cite{Diehl1} demonstrates that the two-defect interaction potential \emph{in the absence of non-mutual interaction} changes sign for $L_d\sim e^{2\bar{K}/\lambda}$ implying that defects unbind beyond this scale. In chiral active hexatics, this is complicated by the non-mutuality of interaction but we find that the \emph{sign} of the interaction between a $\pm 1/6$ pair changes beyond a critical distance within a perturbative treatment to $\mathcal{O}(\lambda^2)$. However, non-mutual interactions between defects complicates the many-body physics, and opens up possibilities for novel behaviours, such as charge-separation, which we will explore in a future publication.

We now discuss how a chiral hexatic phase may arise, in a small enough system, from the melting of a chiral solid. In the crystalline phase, which breaks both rotation and translation symmetries, the bond-angle field is slaved to the displacement field, which we denote by ${\bf u}({\bf x}, t)$, as
\begin{equation}
\epsilon_{ij}\theta=\frac{1}{2}(\partial_i u_j-\partial_j u_i)=W^a_{ij},
\end{equation}
where ${\bsf W}^a$ is the antisymmetric part of the displacement gradient tensor which denotes a rotation of the crystal structure.
Inserting this form of $\theta$ into the active forces $-\zeta\hat{z}\times\nabla\theta$ and $-\zeta_c\hat{z}\times(\hat{z}\times\nabla)\theta$ in \eqref{velhex}, we find that these correspond to the active forces $\zeta\nabla\cdot{\bsf W}^a$ and $-\zeta_c\nabla\cdot(\boldsymbol{\epsilon}\cdot{\bsf W}^a)$ in the solid phase which were ignored in \cite{Deboo2}. We therefore construct a complete theory of active chiral solids, including all active forces in \cite{supp} and formulate a phenomenological theory of dislocation unbinding which yields the hydrodynamic equations of the chiral active hexatic that we have discussed in this paper. This connects the phenomenological parameters in \eqref{bang} and \eqref{velhex} with those in the theory of the solid. 

We close by discussing the generality of our results and its experimental implications. First, though we considered a chiral hexatic on a substrate, our primary results, namely giant number fluctuations in a non-rotating hexatic, waves in the presence of a density-dependent rotation rate and the destruction of quasi-long-range order due to nonlinearities all remain valid even for a hexatic suspension of active particles in an incompressible momentum conserved fluid with $\nabla\cdot{\bf v}=0$ \cite{supp}. 
Next, though our results are strictly applicable for an inherently chiral system, they may have relevance for microscopically \emph{achiral} systems, in which chiral symmetry is broken spontaneously. \AMR{This may have been observed in \cite{palacci}, perhaps due to non-reciprocal interactions between an internal orientational degree of freedom and the local structural organisation \cite{Ano_nonreci, Vincenzo_nonreci}}. 
This may lead to a spontaneous chiral symmetry-broken state in which the structure rotates at a constant rate \cite{Ano_Cesare} which should be described by our theory. We also discuss the potential implication of the nonlinear destabilisation uncovered here for chiral phases with lower angular symmetry such as nematics in the supplement \cite{supp}.
\AMR{Further, while the discussion in this paper was centred around hexatic liquid crystals, the equations we used had an extra symmetry under \emph{independent} rotations of space and spin which would be spoiled only if we expanded the active and passive forces to \emph{fifth} order in gradients. Because of this extra symmetry, the equations of motion considered here are also applicable to a system that has the spatial symmetry of a chiral variant of an XY model -- the compact KPZ equation (distinct from a motile \emph{polar} system \cite{Toner_Tu} which is \emph{not} invariant under independent rotations of space and spin) -- coupled to a conserved quantity. In fact, the Malthusian \cite{Malthusian} version of our theory, Eq. \eqref{comKPZ}, has been used to describe driven dissipative condensates \cite{Diehl1, TonerPRX}. We belive our conclusions regarding non-mutual interaction between defects, which generalises the discussion in \cite{Diehl1}, should be observable in those systems.}
Our results also have implications for multiple experimental systems such as active, chiral vortex lattice phases in motor-microtubule systems \cite{Large-scale vortex lattice emerging from collectively moving microtubules} and in fast-moving bacteria \cite{Fast-Moving Bacteria Self-Organize into Active Two-Dimensional Crystals of Rotating Cells}. Epithelial tissues also often have a hexagonal structure \cite{Hexagonal Packing of Drosophila Wing Epithelial Cells by the Planar Cell Polarity Pathway, Cell surface mechanics and the control of cell shape tissue patterns and morphogenesis, The mechanical anisotropy in a tissue promotes ordering in hexagonal cell packing} and intrinsic chirality have been observed in multiple cellular systems \cite{Spontaneous shear flow in confined cellular nematics}, raising the possibility that certain tissues may be described as active chiral hexatics and our result for giant number fluctuations for non-rotating systems may be tested there at timescales at which cellular birth and death are unimportant. Recently, cells in an isotropically confined epithelial sheet were shown to all spontaneously rotate in the same direction \cite{Ladoux}; our predicted density-orientation wave may be observed in this system.
 Chiral hexagonally ordered phases have also been observed in simulations \cite{Emergent Collective Phenomena in a Mixture of Hard Shapes through Active Rotation, Purely hydrodynamic ordering of rotating disks at a finite Reynolds number, Emergent collective dynamics of hydrodynamically coupled micro-rotors} and our predictions regarding the hexatic phase should be verifiable there.

\begin{acknowledgments}A.M. acknowledges illuminating discussions with Debarghya Bannerjee, Cesare Nardini, Sriram Ramaswamy and Vincenzo Vitelli. M. L. was supported by Marie 
Curie Integration Grant PCIG12-GA-2012-334053, ‘‘Investissements 
d’Avenir’’ LabEx PALM (ANR-10-LABX-0039-PALM), ANR grant 
ANR-15-CE13-0004-03 and ERC Starting Grant 677532. ML’s group belongs to 
the CNRS consortium CellTiss. R. V. was supported by ANR grant PHYMAX and POLCAM.
\end{acknowledgments}

\end{document}